# Accelerating Quantitative Susceptibility Mapping using Compressed Sensing and Deep Neural Network

Yang Gao[1], Martijn Cloos[2,3], Feng Liu[1], Stuart Crozier[1], G. Bruce Pike[4], Hongfu Sun[1]*

[1]School of Information Technology and Electrical Engineering, University of Queensland, Brisbane, Australia

[2]Centre for Advanced Imaging, University of Queensland, Brisbane, Australia

[3]ARC Training Centre for Innovation in Biomedical Imaging Technology, The University of Queensland, Brisbane, QLD, Australia

[4]Departments of Radiology and Clinical Neurosciences, University of Calgary, Calgary, Canada

***Correspondence**: Hongfu Sun

**Address**: Room 538, General Purpose South (Building 78),

University of Queensland, St Lucia QLD 4072, Australia

**Email**: hongfu.sun@uq.edu.au

**Sponsors:**

HS acknowledges support from the University of Queensland (UQECR2057605) and the Australian Research Council (DE210101297).

GBP acknowledges support from the Canadian Institutes for Health Research (FDN-143290), the Natural Science and Engineering Research Council (RGPIN-2017-03880), and Campus Alberta Innovates.

MC acknowledges support from the Australian Research Council (FF200100329) and the facilities of the National Imaging Facility at the Centre for Advanced Imaging

**Abbreviations:**

**QSM**, quantitative susceptibility mapping; **ME-GRE**, multi-echo gradient-echo; **SWI**, susceptibility-weighted imaging; **UTE**, ultra-short TE; **FOV**, field of view; **ROI**, region-of-interest; **DGM**, deep grey matter; **DWM**, deep white matter; **RESHARP**, regularization enabled sophisticated harmonic artifact reduction for phase data; **CS**, compressed sensing; **CS$_{PR}$**, CS with periodic regularizations; **CS$_{PC}$**, CS with phase cycling algorithm; **CNN**, convolutional neural network; **AUTOMAP**, automated transform by manifold approximation; **Mag-Unet**, U-net trained on magnitude images; **Phase-Unet**, U-net trained on phase images; **DCRNet**, Deep Complex Residual Network; **BET**, brain extraction tool; **FSL**, FMRIB Software Library; **AF**, accelerating factor; **COSMOS**, Calculation Of Susceptibility through Multiple Orientation Sampling; **GP**, globus pallidus; **PU**, putamen; **CN**, caudate nucleus; **TH**, thalamus; **SN**, substantia nigra; **RN**, red nucleus; **PSNR**, peak signal to noise ratio; **SSIM**, structural similarity; **SSE**, sum of squared errors; **MS**, multiple sclerosis; **ppb**, part-per-billion.

**Abstract**

Quantitative susceptibility mapping (QSM) is an MRI phase-based post-processing method that quantifies tissue magnetic susceptibility distributions. However, QSM acquisitions are relatively slow, even with parallel imaging. Incoherent undersampling and compressed sensing reconstruction techniques have been used to accelerate traditional magnitude-based MRI acquisitions; however, most do not recover the full phase signal due to its non-convex nature. In this study, a learning-based Deep Complex Residual Network (DCRNet) is proposed to recover both the magnitude and phase images from incoherently undersampled data, enabling high acceleration of QSM acquisition. Magnitude, phase, and QSM results from DCRNet were compared with two iterative and one deep learning methods on retrospectively undersampled acquisitions from six healthy volunteers, one intracranial hemorrhage and one multiple sclerosis patients, as well as one prospectively undersampled healthy subject using a 7T scanner. Peak signal to noise ratio (PSNR), structural similarity (SSIM) and region-of-interest susceptibility measurements are reported for numerical comparisons. The proposed DCRNet method substantially reduced artifacts and blurring compared to the other methods and resulted in the highest PSNR and SSIM on the magnitude, phase, local field, and susceptibility maps. It led to 4.0% to 8.8% accuracy improvements in deep grey matter susceptibility than some existing methods, when the acquisition was accelerated four times. The proposed DCRNet also dramatically shortened the reconstruction time by nearly 10 thousand times for each scan, from around 80 hours using conventional approaches to only 30 seconds.

**Introduction**

Quantitative susceptibility mapping (QSM) is an MR phase post-processing technique that has been developed to quantify the magnetic susceptibility distribution of the human brain [1, 2]. It has shown significant clinical potential for studying multiple neurological disorders, including multiple sclerosis [3], Alzheimer's disease [4], Parkinson's disease [5], alcohol use disorders [6], and intracranial hemorrhage [7], as well as healthy aging [8]. QSM scans are generally acquired with a three-dimensional (3D) gradient-recalled-echo (GRE) sequence [9, 10], while other sequences such as ultra-short TE (UTE) [11], multi-echo MP2RAGE [12-14], and water saturation shift referencing (WASSR [15]) have also been proposed. These acquisition methods are relatively slow (usually 5-10 minutes with parallel imaging) since large areas of k-space need to be traversed, with each sampling line accumulating scan time [16, 17]. Gradient-echo EPI sequences [18, 19] can significantly shorten the QSM acquisition time to seconds; however, they come with the compromise of substantially increased image distortion and reduced image resolution.

Incoherent k-space undersampling strategies in combination with compressed sensing (CS) reconstruction techniques have been used to substantially accelerate magnitude-based MRI scans. Since the early work of Lustig et al. [20], an increasing number of approaches, such as sparsity-based CS-MRI models [21-25], Low-Rank models [26-29], and deep learning frameworks [30-38], have been established to recover magnitude images from sparsely sampled k-space data. However, MR images are inherently complex-valued with an equally important phase component that is essential for phase-based contrasts such as QSM. The image sparsity assumption commonly used in previous CS-MRI methods does not apply to the phase since it is periodic and wraps, which makes it an inherently non-convex problem. More advanced reconstruction methods have recently been proposed to reconstruct both magnitude and phase images by extending traditional CS-MRI frameworks. For instance, Zhao et al. [39] designed a phase regularization that is periodic and effective against phase wraps. Ong et al. [40] developed a phase cycling technique invariant to the phase wraps by shifting the wrap positions in each iteration; this was successfully applied to accelerate mouse brain QSM in a recent work [17]. However, these phase regularization methods lead to blurring in the reconstruction results, and the algorithms are computationally expensive and slow, which makes them impractical for high-resolution 3D MRI volumes.

Deep learning has received increasing attention in various imaging research areas [30, 41] as an alternative to conventional iterative algorithms. One method, named Automated Transform by Manifold Approximation (AUTOMAP) [32], can directly recover the magnitude and phase maps from the k-space signals (sensor-domain data). However, this network only works for small-size images due to the memory-consuming fully connected layers. Another study [34] proposed two separate convolutional neural networks (CNN) for magnitude and phase reconstruction, respectively. However, the phase wraps in the reconstructed images are disrupted, making it problematic for quantitative phase imaging applications. Furthermore, these methods only showed limited results on 2D phase reconstructions, and their feasibility and robustness on the QSM reconstruction pipeline have not been investigated.

In this study, a Deep Complex Residual Network (DCRNet) is designed, by incorporating the complex convolutions [42-44] into a deep residual network backbone to recover both MR magnitude and quantitative phase images, thus enabling the acceleration of QSM acquisitions. DCRNet processes the real and imaginary images with complex convolutions inspired by the multiplication of complex numbers, which may better exploit the correlation between real and imaginary parts of a complex image [43]. Qualitative and quantitative comparisons are performed among different iterative and deep learning algorithms on the magnitude, phase, and QSM images from healthy subjects and patients with intracranial hemorrhage and multiple sclerosis. The undersamplings are performed both retrospectively on existing 3T datasets and prospectively at 7T. The performance of the proposed DCRNet under different accelerating factors (i.e., 2×, 4×, 6×, 8×) is also investigated.

**Theory**

**General Forward Model**

The general forward model of CS undersampled MRI [39, 40] can be formulated as:

$$\bm{y} = \bm{AF}(\bm{m} \odot \exp(i\bm{\varphi})), \qquad (1)$$

where $i = \sqrt{-1}$; $\bm{F}$ is the discrete Fourier transform matrix; $\bm{m}$ and $\bm{\varphi}$ represent the magnitude and phase images to be reconstructed; $\odot$ represents the elementwise multiplication; $\exp(\cdot)$ denotes the element-wise exponential operator; $\bm{A}$ is the CS undersampling matrix from a designed subsampling mask; $\bm{y}$ is the CS undersampled k-space data.

**Conventional Method for Magnitude-Only Reconstruction**

In the conventional CS reconstruction method for MR magnitude images [20], a sparsifying transformation regularization (e.g., total variation, wavelet transform) is usually incorporated into the cost function to derive an optimized solution:

$$\operatorname*{argmin}_{\boldsymbol{m}} \frac{1}{2}\|\boldsymbol{y} - \boldsymbol{AF}(\boldsymbol{m} \odot \exp(i\boldsymbol{\varphi}))\|^2 + \lambda \Re(\boldsymbol{m}), \qquad (2)$$

where $\|\cdot\|^2$ is the vector $l2$ norm; $\Re(\cdot)$ represents the sparsifying regularization function; $\lambda$ is the regularization parameter. In this scheme, the phase signal is estimated from the fully-sampled low-frequency k-space region, which can be used to correct low-order phase variation and better sparsify the MR magnitude image be reconstructed [20]. Although this method can obtain fairly robust magnitude images and has been widely adopted for accelerating magnitude-based scans, it is not feasible to restore high-resolution GRE phase images for QSM.

**Iterative Approaches with Phase Regularizations**

To reconstruct both magnitude and phase, some previous methods [39, 40] added magnitude and phase regularizations, respectively:

$$\operatorname*{argmin}_{\boldsymbol{m},\boldsymbol{\varphi}} \frac{1}{2}\|\boldsymbol{y} - \boldsymbol{AF}(\boldsymbol{m} \odot \exp(i\boldsymbol{\varphi}))\|^2 + \lambda_1 \Re_m(\boldsymbol{m}) + \lambda_2 \Re_\varphi(\boldsymbol{\varphi}), \qquad (3)$$

where $\|\cdot\|^2$ is the vector $l2$ norm; $\Re_m$ and $\Re_\varphi$ are regularization functions for magnitude and phase, respectively; $\lambda_1$ and $\lambda_2$ are empirical regularization parameters. However, it is reported [39, 40] that using conventional CS regularizers (e.g., wavelet transform and finite differencing matrix) on the phase component leads to significant reconstruction errors and artifacts at phase wraps because the phase image $\varphi$ appears as an exponential term in the cost function, making the optimization process inherently non-convex.

Zhao et al. [39] suggested using periodic regularizations in the complex exponential form to solve this problem:

$$\operatorname*{argmin}_{\boldsymbol{m},\boldsymbol{\varphi}} \frac{1}{2}\|\boldsymbol{y} - \boldsymbol{AF}(\boldsymbol{m} \odot \exp(i\boldsymbol{\varphi}))\|^2 + \lambda_1 \|\boldsymbol{Wm}\|_1 + \lambda_2 \sum_{k=1}^{K} \psi(|[\boldsymbol{C}\exp(i\boldsymbol{\varphi})]_k|), \qquad (4)$$

where $\|\cdot\|_1$ is the vector $l1$ norm operator; $\boldsymbol{W}$ is the wavelet transform matrix; $\boldsymbol{C}$ is the finite differencing matrix; $K$ is the number of rows of $\boldsymbol{C}$; $\psi(x) = \delta^2(\sqrt{1 + \left|\frac{x}{\delta}\right|^2} - 1)$ is an edge-preserving potential function, and $\delta$ (empirically set as 0.005) is the parameter to tune the edge-preserving weight. However, this method is not general and cannot be applied to phase-based water-fat or flow imaging [40]. Alternatively, a more generalized algorithm [40] named phase cycling was developed, which is equivalent to an inexact proximal gradient method applied to the following cost function:

$$\underset{\boldsymbol{m},\boldsymbol{\varphi}}{\mathrm{argmin}} \frac{1}{2}\|\boldsymbol{y} - \boldsymbol{AF}(\boldsymbol{m} \odot exp(i\boldsymbol{\varphi}))\|_2^2 + \lambda_1 \|\boldsymbol{Wm}\|_1 + \lambda_2 \frac{1}{|\mathcal{P}|}\sum_{\boldsymbol{p}\in\mathcal{P}}\|\boldsymbol{W}(\boldsymbol{\varphi}+\boldsymbol{p})\|_1, \quad (5)$$

where $\mathcal{P}$ is a collection of phase-shifting matrices generated from the zero-filling reconstruction to shift the phase wraps. The phase cycling technique achieves the reconstruction robustness against the phase wraps by randomly shifting the phase wraps in each iteration and effectively averaging the artifacts spatially to a low level. This algorithm has been applied to accelerate mouse brain QSM in previous work [17]; however, it is slow (i.e., around 40 seconds for a single 2D slice) and has not been investigated using high-resolution 3D human brain QSM data.

In this study, the above two iterative methods (Eqs. (4) and (5)) referred to as $CS_{PR}$ (periodic regularizations) and $CS_{PC}$ (phase cycling) were implemented and compared with the proposed new method, detailed below.

**Methods**

**QSM Accelerating Framework**

The proposed deep learning-based QSM accelerating framework is shown in Fig. 1, which consists of two parts: (i) the MRI magnitude and quantitative phase reconstruction from the CS undersampled k-space data using the proposed DCRNet, (ii) the QSM processing pipeline, including brain mask generation with BET [45] from the FMRIB Software Library (FSL) [46], phase unwrapping with a best path method [47], magnitude-weighted fitting of multi-echo phase with echo times, background field removal with RESHARP [48], and finally dipole inversion (single-orientation: xQSM [49], multiple orientations: Calculation Of Susceptibility through Multiple Orientation Sampling (COSMOS) [50]).

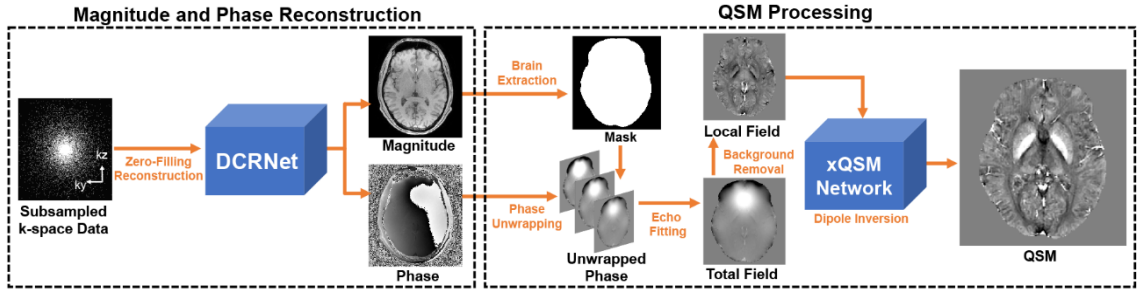

**Figure 1**. Overview of the proposed QSM accelerating scheme. The left part demonstrates the DCRNet-based MRI magnitude and quantitative phase image reconstruction, while the right part shows the established QSM reconstruction pipeline, including the learning-based xQSM for dipole inversion.

**DCRNet for Complex-valued MRI Reconstruction**

The proposed DCRNet architecture is shown in Fig. 2(a), specifically designed for complex-valued MR images, adding complex-valued operations to a deep residual network backbone. Although QSM is reconstructed from 3D phase images, the proposed network is a 2D residual neural network to save GPU memory. It is straightforward to apply this 2D network to 3D MRI reconstruction in a slice-by-slice manner. As shown at the top of Fig. 2(a), the 3D subsampled k-space data are first converted into 2D k-space slices after a 1D Fourier Transform along the fully-sampled $k_r$ (frequency-encoding) direction. The 2D k-space slices are also used for data consistency (details below). The proposed DCRNet is a fully convolutional neural network, enabling it to process complex images of any size. The details of each building block, i.e., the input layer, residual blocks, the output layer, and the data consistency layer, are described as follows:

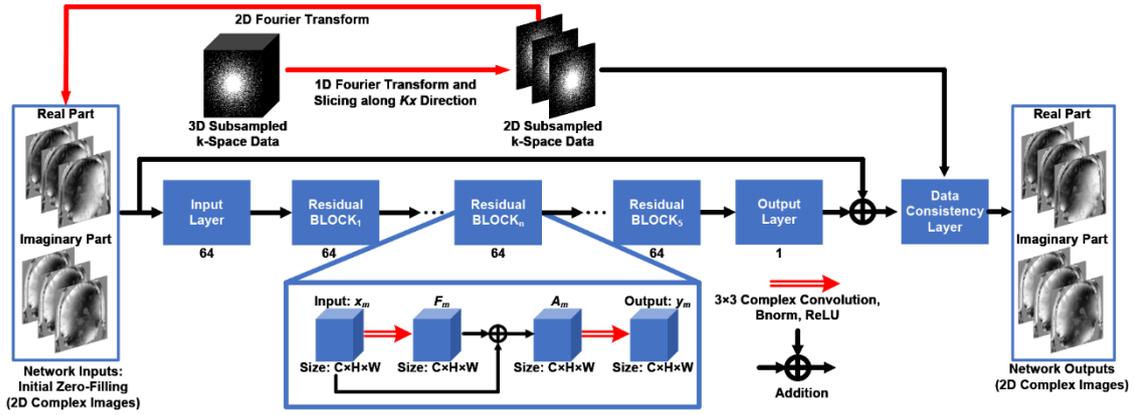

(a) DCRNet Architecture

(b) Complex Convolution

**Figure 2**. The proposed DCRNet, which is developed from a deep residual network backbone using complex convolutional operations. (a) demonstrates the deep residual network backbone diagram, while (b) illustrates the proposed complex convolution operation, which is developed from the complex multiplication principle.

First, an input layer, consisting of a complex convolution followed by a batch-normalization operation and an activation function, is used to extract the high-dimensional features from the complex input $x_0$ (i.e., the initial zero-filling reconstruction):

$$y_0 = ReLU(BN(W_0 * x_0 + b_0)), \qquad (6)$$

where $W_0 \in \mathbb{C}^{64\times1\times3\times3}$ and $b_0 \in \mathbb{C}^{64\times1}$ are the complex convolutional kernel and bias of this input layer, and $\mathbb{C}$ is the complex number set; $BN$ and $ReLU$ represent the conventional batch-normalization operation and the Rectified Linear Unit activation function, which are applied separately to the real and imaginary components of the complex feature maps; $*$ is the complex convolution operation (detailed in the following section); $y_0$ represents the output features of this layer.

The DCRNet then cascades the middle residual block five times, with each block consisting of two complex convolutions and one addition:

$$F_m = ReLU(BN(W_m^0 * x_m + b_m^0))$$
$$A_m = F_m + x_m$$
$$y_m = ReLU(BN(W_m^1 * A_m + b_m^1))$$
$$x_{m+1} = y_m$$
(7)

where $x_m$ and $y_m$ are the input and output of the $m \in \{1, 2, \ldots, 5\}$ residual block, with $x_1 = y_0$; $W_m^0 \in \mathbb{C}^{64 \times 64 \times 3 \times 3}$ and $W_m^1 \in \mathbb{C}^{64 \times 64 \times 3 \times 3}$ are the complex convolutional kernels, and $b_m^0 \in \mathbb{C}^{64 \times 1}$ and $b_m^1 \in \mathbb{C}^{64 \times 1}$ are the bias terms; $F_m$ and $A_m$ are intermediate feature maps.

An output layer is applied to generate an image of the same size as the input by aggregating the high-dimensional features from previous blocks. Additionally, a skip connection between the input and result of this output layer is added, forming a whole residual block, which helps network training converge faster [51] and mitigates the gradient vanishing problem [52]. The corresponding output image $y_6$ can be formulated as:

$$y_6 = x_0 + W_6 * x_6 + b_6, \qquad (8)$$

where $x_6 = y_5$ is the output of the 5$^{th}$ middle residual block and $W_6 \in \mathbb{C}^{1 \times 64 \times 3 \times 3}$ and $b_6 \in \mathbb{C}^{1 \times 1}$ correspond to the complex convolutional kernel and bias term in the output layer, respectively.

Finally, a data consistency layer is adopted at the end of the DCRNet to generate the final reconstruction image $y_{rec}$ from $y_6$ and the network input $x_0$, which can be expressed as:

$$Y_{rec}(k) = \begin{cases} Y_6(k) & , if\ k \notin \Omega \\ \frac{\lambda X_0(k) + Y_6(k)}{(1+\lambda)}, & if\ k \in \Omega \end{cases} \qquad (9)$$

where $Y_6(k)$ and $X_0(k)$ are the corresponding k-space representations of $y_6$ and $x_0$; $\Omega$ is the collection of the sampling positions determined by the subsampling masks; $\lambda$ is a weighting parameter depending on the noise levels [36, 43] and is set as learnable in the current work. The final reconstruction of the DCRNet, $y_{rec}$, is obtained from the inverse Fourier Transform of the k-space signal $Y_{rec}(k)$.

**Complex Convolutional Operations**

In this study, the complex convolutional layer is developed based on the multiplication rules of complex numbers [43, 44]. The diagram of the proposed complex convolution is shown in Fig. 2(b). The complex convolution between a complex input $X = X_R + i X_I$ and a complex convolutional kernel $W = W_R + iW_I$ is represented as:

$$Y = X \circledast W,$$

$$\text{where } \begin{cases} Y = Y_R + i Y_I \\ Y_R = X_R * W_R + X_I * W_I \\ Y_I = X_R * W_I + X_I * W_R \end{cases} \quad (10)$$

Here $\circledast$ represents the complex convolution and $*$ represents the conventional real-valued convolution operation in traditional CNNs; in summary, a complex convolution can be split into four traditional real-valued complex convolutions and two addition operations.

Other conventional operations, such as the batch-normalization and *ReLU* function on a complex feature map $X = X_R + i X_I$ are separately applied to the real and imaginary components:

$$\begin{aligned} BN(X) &= BN(X_R) + i \cdot BN(X_I) \\ ReLU(X) &= ReLU(X_R) + i \cdot ReLU(X_I) \end{aligned} \quad (11)$$

**In vivo experiments and Data Preparation**

Whole-head 3D multi-echo GRE (ME-GRE) raw k-space data (8 unipolar echoes, first TE: 3 ms, echo spacing: 3.3 ms, TR 29.8 ms, FOV: 256×256×128 mm³, voxel size: 1mm isotropic, 12-channel head coil with ASSET acceleration factor of 2, 3/4 partial Fourier in both phase and slice directions) from ten healthy volunteers were acquired at 3T (Discovery 750, GE Healthcare, Milwaukee, WI) and used to generate training labels. The coil-combined magnitude and phase images exported from the scanner were transformed into the Fourier domain to simulate the fully-sampled k-space data of these scanner reconstructed images.

These simulated single-channel 3D fully-sampled k-space datasets were retrospectively subsampled using a 2-D variable density k-space sampling mask in the $k_y$-$k_z$ plane, generated with the probability density function [53] formulated as:

$$PDF = \exp(-P_a \times (\sqrt{\frac{k_y^2}{n_y} + \frac{k_z^2}{n_z}})^{P_b}), \quad (12)$$

where $P_a$ and $P_b$ are two experimental parameters, $[n_y, n_z]$ is the k-space matrix size, and $k_y$, $k_z$ are the k-space coordinates. Fig. 3 (a) shows the undersampling masks generated for training of the proposed networks ($P_a$=7, $P_b$=1.8 for 2×, $P_a$=12, $P_b$=1.8 for 4×, $P_a$=17, $P_b$=1.8 for 6×, and $P_a$=22, $P_b$=1.8 for 8× accelerations). The undersampling aliasing artifacts from the zero-filling reconstructions are demonstrated in Fig. 3 (b) in three orthogonal views in both magnitude and phase images. These initial zero-filling results, reformatted as real and imaginary images, are used as the training inputs for DCRNet.

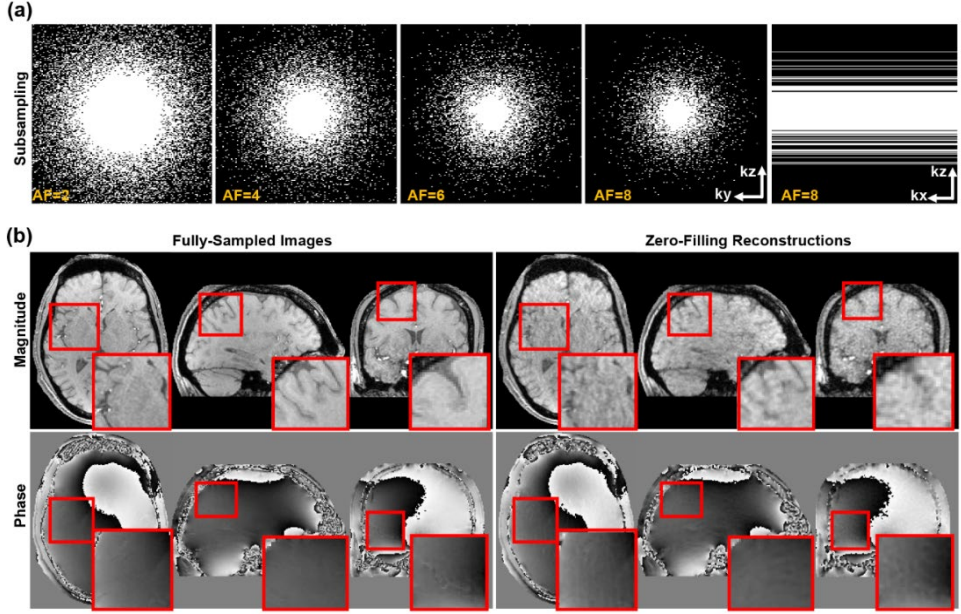

**Figure 3.** (a) shows the subsampling masks of different acceleration factors (AFs). Each dot in the $k_y$-$k_z$ plane (the first 4 masks) represents a fully-sampled readout line in the the $k_x$ direction, as shown in the $k_x$-$k_z$ plane (the last mask). (b) illustrates the aliasing artifacts from zero-filling reconstructions in three orthogonal views in both magnitude and phase images from AF = 4.

Next, these 3D zero-filling image volumes were sliced into 2D complex images along the $k_x$ direction and were normalized (i.e., divided by the maximum intensity of the 3D complex image volume) before network training. A total of 16800 complex GRE images (size: 256×128) retrospectively undersampled from 10 healthy subjects were used for network training.

To compare the performance of the proposed DCRNet with other methods, another fourteen fully-sampled 3D ME-GRE MRI scans from six healthy volunteers were acquired and undersampled retrospectively using the subsampling mask at AF = 4. Two of these subjects were scanned at five different head orientations, while another four subjects were obtained with a single head orientation. Two more ME-GRE datasets, one intracranial hemorrhage and one multiple sclerosis patients, were also fully-sampled and retrospectively undersampled to validate the performance of the proposed DCRNet in the presence of pathology.

To implement and test the proposed QSM acceleration framework, including both acquisition and reconstruction, one healthy volunteer was scanned with a modified GRE sequence at 7T (Magnetom Classic, Siemens Healthcare, Erlangen, Germany). The acquisition was prospectively undersampled by AF of 4, using the pre-designed sampling mask shown in Fig. 3(a). A single transmit 32 channel head coil (Nova Medical, Wilmington, MA, USA) was used for signal reception. The GRE scan parameters were as follows: 8 unipolar echoes, first TE 3.4 ms, echo spacing 3.5 ms, TR 31 ms, BW 480 Hz/pixel, FOV 256×192×128, voxel size 1mm isotropic, scan time 3 mins 10 seconds. Zero-filling reconstructed was first performed for each receiver channel. Coil sensitivity maps were estimated using the POEM method before properly combined for phase [13]. For the DCRNet method, the combined zero-filling images were fed into the deep neural network without the data consistency module. For the $CS_{PC}$ reconstruction method, raw k-space measurements from individual channels and estimated coil sensitivity maps were used according to the original paper [40].

**Network Training**

The mean squared error loss function was adopted to train the proposed DCRNet. An extra noise adding layer detailed in one recent work [49] was applied to improve the network's robustness against noise. All trainable parameters were initialized with normally distributed random numbers with a mean of zero and a standard deviation of 0.01 and optimized using the adaptive moment estimation (Adam optimizer [54]) with a mini-batch size of 32. The networks were trained for 100 epochs, and the learning rate was set to $10^{-3}$, $10^{-4}$, and $10^{-5}$ for the first 40 epochs, 40-80 epochs, and the final 20 epochs, respectively. It took around 8 hours for network training on 2 Nvidia Tesla V100 GPUs (32 GB). The networks were implemented using Pytorch 1.8, and the source codes and trained networks are available at https://github.com/YangGaoUQ/DCRNet.

**Quantitative Performance Evaluation**

The proposed DCRNet was compared with two iterative algorithms (i.e., $CS_{PR}$ [39] and $CS_{PC}$ [40]) and one deep learning method with separate magnitude and phase Unets [34] (referred to as Mag-Unet and Phase-Unet) at an accelerating factor (AF) of 4×. The Mag-Unet and Phase-Unet were re-trained using the same datasets and optimizer settings as for DCRNet. The reconstructions of deep learning methods were computed on a high-performance computing cluster using one Tesla V100 GPU (32 GB), while the reconstructions of the iterative methods were accomplished with 5 Intel Xeon E5-2680V3 2.5Ghz CPU (12 core processors) on a high-performance computing cluster.

Peak signal-to-noise ratio (PSNR), structural similarity (SSIM), and susceptibility measurements of one deep white matter (DWM) region and six deep grey matter (DGM) regions (i.e., globus pallidus (GP), putamen (PU), caudate nucleus (CN), thalamus (TH), substantia nigra (SN), red nucleus (RN)) were measured for quantitative comparison using data from the 6 healthy volunteers.

The hemorrhage susceptibility measurements were reported for the intracranial hemorrhage patient, and voxel-wise linear regressions were carried out to compare different QSM accelerating methods. Visual inspection was performed to assess the QSM image quality and the delineation of white matter lesions in the multiple sclerosis patient.

Different subsampling reconstruction methods were applied to restore the full magnitude, phase, and QSM images from the prospectively undersampled subject. The degree of aliasing artifacts and the effects on image SNR and resolution were evaluated and compared among different methods.

The performance of the proposed DCRNet with varying AFs (2×, 4×, 6×, 8× times) was also investigated on one healthy subject.

## Results

**QSM Undersampling Reconstruction for Healthy Subjects**

The magnitude, phase, local field (by RESHARP [48]), and susceptibility maps (by xQSM [49]) reconstructed at 4× AF from different methods was compared on one healthy volunteer, as shown in Fig. 4. All image contrasts show that the proposed DCRNet led to the best PSNR and SSIM (e.g., 41.62/0.97 for QSM reconstructions). $CS_{PR}$ and Phase-Unet exhibited substantial artifacts (red arrows) in local field maps and QSM images near the strong susceptibility sources (e.g., sinuses), which were absent in other methods. The reconstruction time of one brain volume (size: 256×256×128×8) for the deep learning methods (i.e., around 30 seconds for DCRNet and 70 seconds for Phase-Unet) is substantially shorter than the traditional iterative methods (around 21 and 28 hours for $CS_{PC}$ and $CS_{PR}$, respectively).

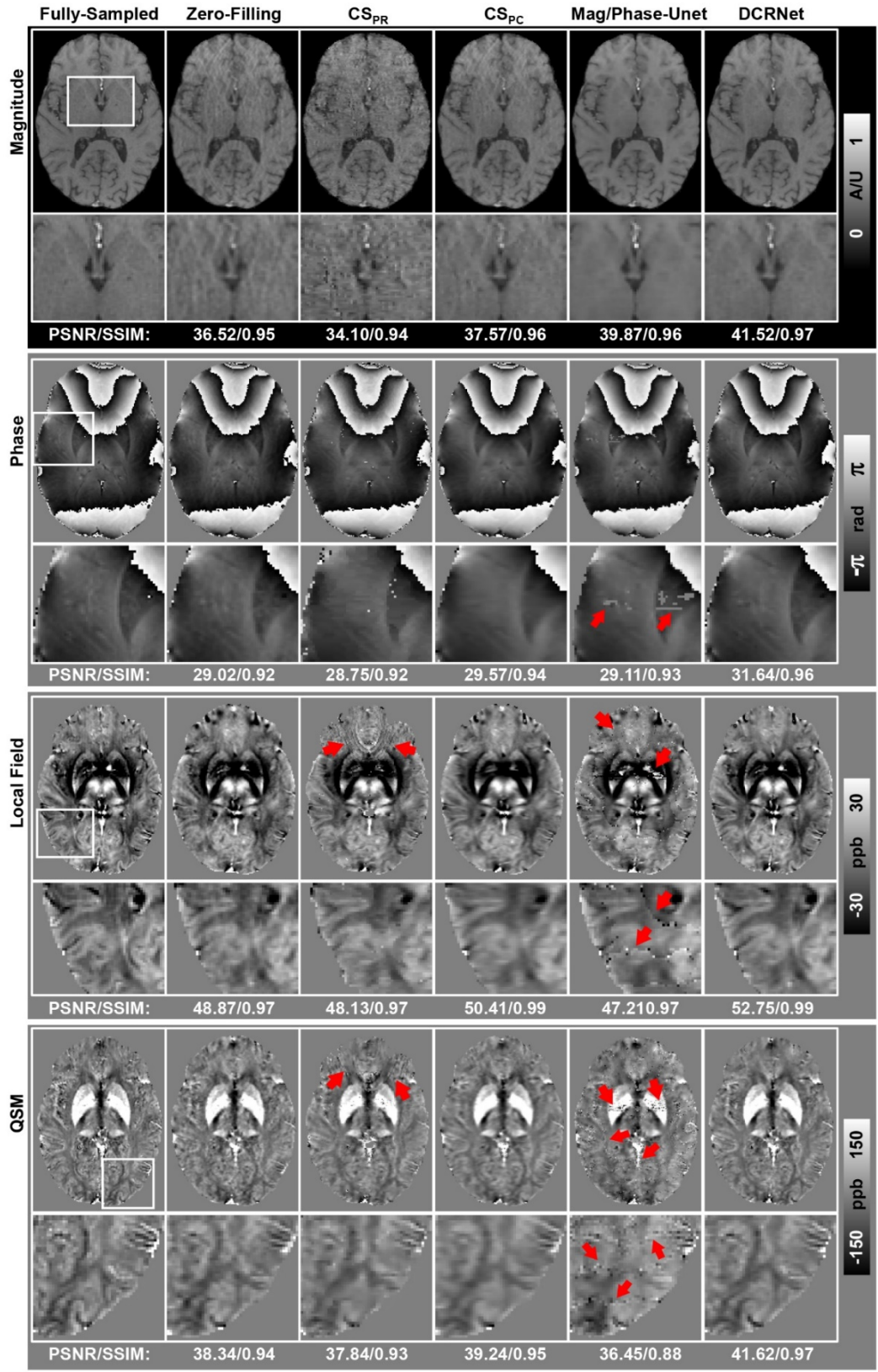

**Figure 4**. Comparison of different MRI phase accelerating methods (at 4× AF) on one healthy subject acquired with a single head orientation. The top four rows compare the reconstructions of magnitude and phase images, while the bottom four rows show the local field and QSM reconstructions from different methods. PSNR and SSIM relative to the ground truth (fully-sampled data) are reported under the images. Red arrows point to the overwhelming artifacts introduced in CS$_{PR}$ and Phase-Unet methods.

Susceptibility measurements of one DWM and six DGM regions using different reconstruction methods from a total of 14 scans were compared in Fig. 5. Paired t-test found that all methods except DCRNet showed significant difference relative to the fully-sampled ground truth in GP, with average QSM deviations of 5.1% (*P* = 0.02), 6.7% (*P* = 0.03), 8.0% (*P* = 0.04), 9.9% (*P* = 0.01), and 1.1% (*P* = 0.08) in zero-filling, CS$_{PR}$, CS$_{PC}$, Phase-Unet, and DCRNet, respectively. Furthermore, the Phase-Unet reconstructions showed the largest standard deviations in all region-of-interest (ROI) measurements (e.g. 123% and 155% larger than other methods in RN and DWM) due to the overwhelming artifacts from corrupted phase wraps recovery.

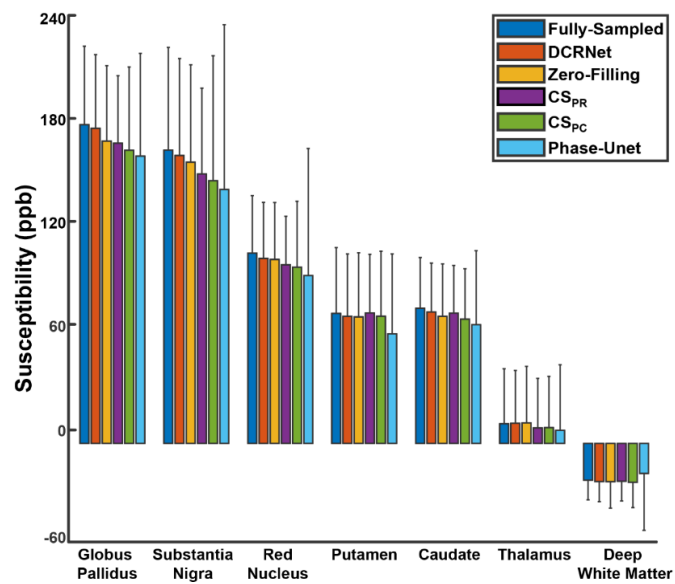

**Figure 5**. One deep white matter and six deep grey matter susceptibility measurements (in mean and standard deviation) from different QSM accelerating methods at 4× AF using 14 scans from 6 healthy subjects.

The COSMOS results (at 4× AF) from DCRNet with zero-filling and CS$_{PC}$ methods on two healthy subjects (five head orientations acquired for each) were compared in Fig. 6. Overall, the proposed deep residual network DCRNet led to higher PSNR and SSIM (47.57/0.99) than CS$_{PC}$ (44.75/0.97) and zero-filling reconstruction (44.54/0.97). Significant image smoothing was found in the CS$_{PC}$ method, while the DCRNet preserved fine details as seen in the zoomed-in regions in Fig. 6. Substantial artifacts near the cortex were observed in zero-filling results (red arrows). According to the table at the bottom of Fig. 6, the CS$_{PC}$ method exhibited the highest susceptibility deviation from fully-sampled reconstruction in GP (5.1%), SN (10.3%), and RN (23.9%) regions. In contrast, the smallest deviations were found in the DCRNet results for all seven brain regions measured.

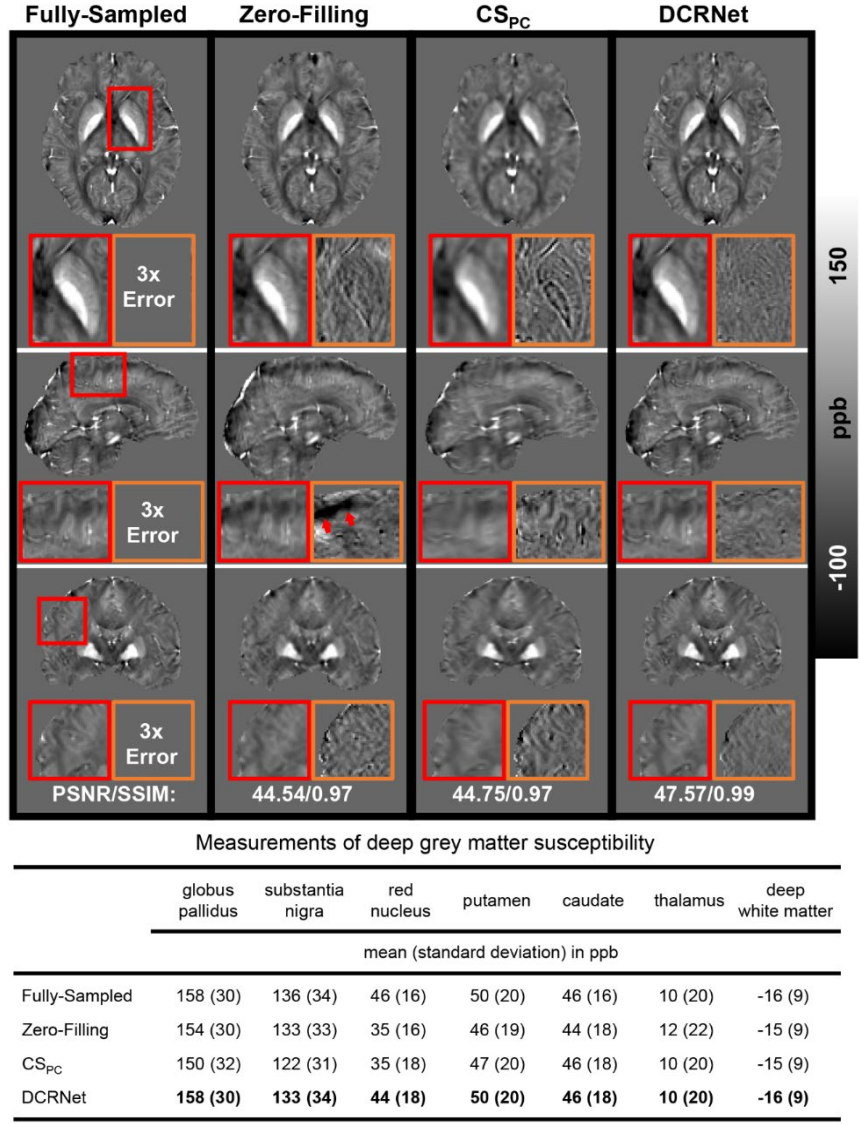

**Figure 6**. Comparison of the proposed DCRNet with zero-filling and $CS_{PC}$ methods on COSMOS results at 4× AF from two healthy volunteers, each acquired in 5 different head positions. COSMOS maps, zoomed-in regions and error maps from one subject were shown in three orthogonal views, with PSNR and SSIM relative to the fully-sampled ground truth displayed below the images. Red arrows point to the apparent artifact in the error map. The table reports the susceptibility measurements (mean and standard deviation) in one DWM region and six DGM regions from the two COSMOS subjects. The best measurements closest to full-sampled are highlighted in bold.

## QSM Undersampling Reconstruction in the Presence of Pathology

Accelerated QSM acquisition results from the intracranial hemorrhage patient (at 4× AF) were compared between the proposed DCRNet and $CS_{PC}$ methods in Fig. 7. QSM from DCRNet appeared similar to the fully-sampled result from visual inspection, with the highest PNSR and SSIM (36.04/0.97). In contrast, zero-filling results showed severe artifacts and $CS_{PC}$ exhibited over-smoothness. A hemorrhage ROI was drawn on a sagittal slice (blue contour) in Fig. 7(a), and the susceptibility of each voxel was compared with ground truth through linear regression in Fig. 7(b)-(d). It confirmed that the proposed DCRNet ($R^2$: 0.65, SSE: 1.89) led to the most accurate hemorrhage susceptibility measurements, compared with zero-filling ($R^2$: 0.52, SSE: 3.12) and $CS_{PC}$ ($R^2$: 0.46, SSE: 2.28) methods.

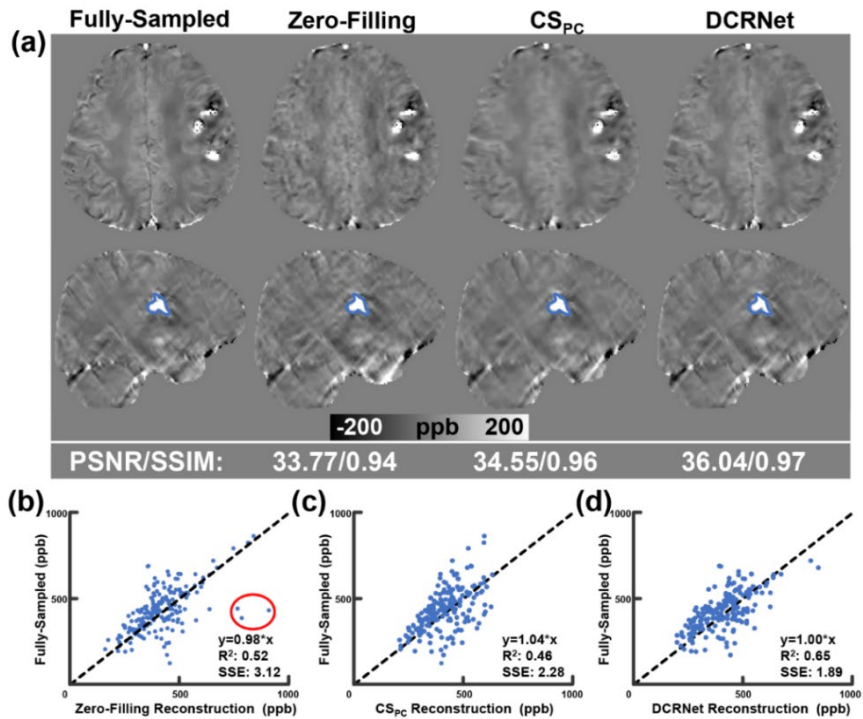

**Figure 7**. Comparison of the proposed DCRNet and $CS_{PC}$ (at 4× AF) on the subsampled MRI data from an intracranial hemorrhage patient. QSM images are illustrated in (a) along with PSNR and SSIM reported underneath. (b)-(d) demonstrate the scatter plots of susceptibility values in the brain lesion (blue contour in (a)) of different reconstruction methods (x-axis) against the ground truth (y-axis). Linear regression results are reported in the bottom-right corner for each plot. The red circle in (b) highlights some of the hemorrhage measurements in zero-filling results highly deviated from the fully-sampled references.

The proposed DCRNet and $CS_{PC}$ QSM accelerating methods were applied to the multiple sclerosis patient in Fig. 8. All QSM acceleration reconstruction methods successfully detected the brain lesions, as indicated by the red arrows. However, the zero-filling reconstruction method suffered from significant artifacts, and the $CS_{PC}$ method led to an apparent over-smoothing effect compared to DCRNet and full-sampled QSM results.

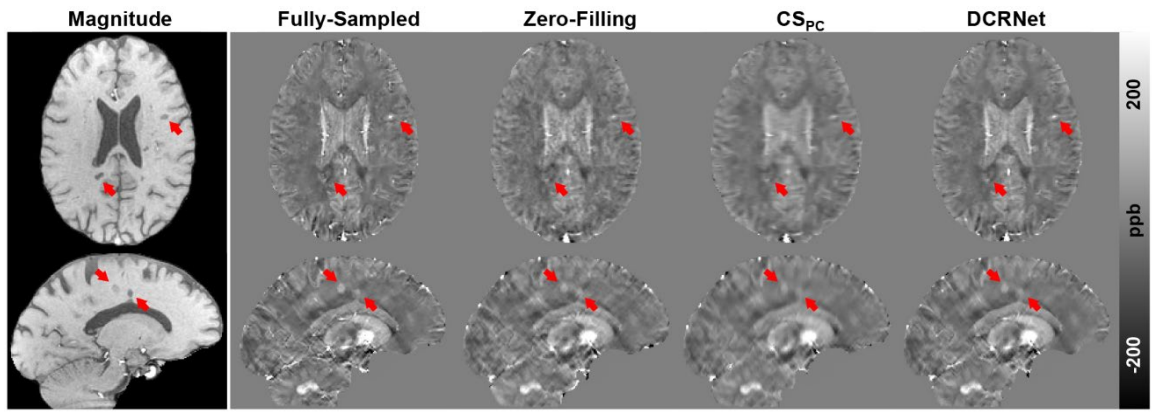

**Figure 8**. Comparison of different QSM accelerating reconstruction methods (at 4× AF) on one multiple sclerosis (MS) patient. The first column shows fully-sampled GRE magnitude images in two orthogonal slices, while the corresponding QSM results from different methods are illustrated from the second column. Red arrows point to the MS lesions that are identified by visual inspection.

**QSM Reconstruction for the Prospectively Undersampled Acquisition**

The zero-filling, $CS_{PC}$, and DCRNet methods for the prospectively undersampled subject were compared in Fig. 9. The in vivo acquisition was performed with a subsampled GRE sequence (AF = 4) at a 7T. The coil-combined magnitude images from the zero-filling method displayed apparent aliasing artifacts similar to retrospectively undersampled results. The $CS_{PC}$ method exhibited residual artifacts in the magnitude reconstructions, while by contrast, the aliasing artifacts were removed after DCRNet reconstruction. Local field and susceptibility maps from zero-filling showed poor SNR, significantly degraded the image qualities. While the $CS_{PC}$ and DCRNet methods substantially decreased the noise level and increased SNR in the local field and susceptibility maps, as evident in the zoomed-in images. The reconstruction time of this in vivo subject (matrix size: 256×192×128×8(TEs)×32(receivers)) for the DCRNet (30 seconds) is almost 10 thousand times faster than the $CS_{PC}$ method (about 80 hours).

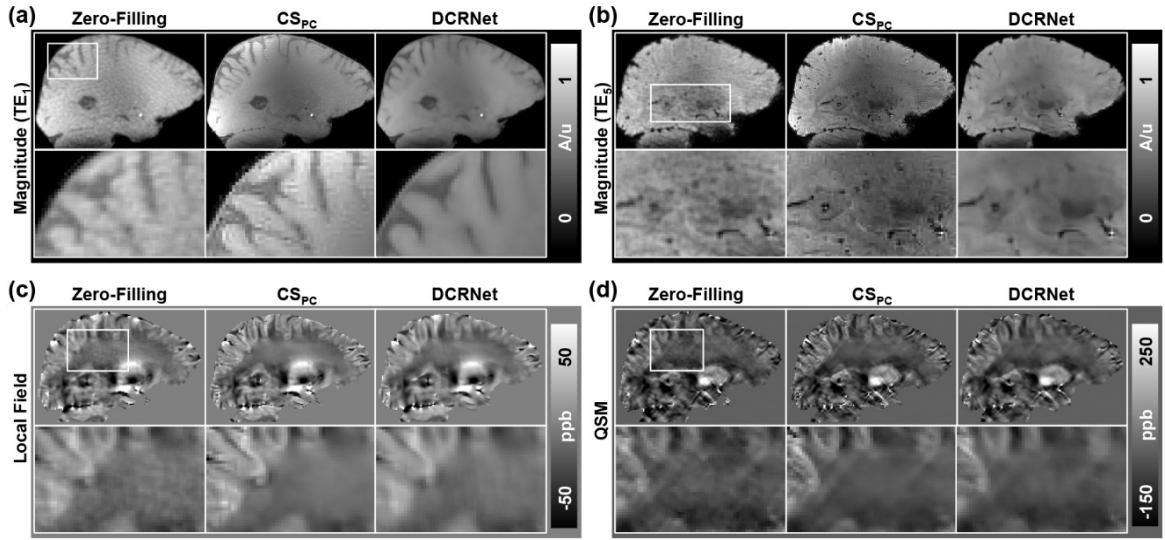

**Figure 9**. Comparison of different reconstruction methods on one subject, prospectively undersampled at 4× AF. (a, b) reconstructed magnitude images at two different TEs (TE$_1$= 3.4 ms, TE$_5$ = 17.4 ms); (c) local field maps; (d) QSM results from zero-filling, CS$_{PC}$, and DCRNet methods. White box regions of various tissue contrasts and SNR levels are zoomed-in at the bottom rows in (a)-(d).

**Evaluation with Different Accelerating Factors**

The reconstruction results of the proposed DCRNet under different AFs (2×, 4×, 6×, and 8×) were demonstrated in Fig. 10 on one healthy subject, undersampled retrospectively. None of the images showed visually apparent artifacts, and the overall QSM contrast was adequately preserved in all AFs. However, the numerical metrics (PSNR and SSIM) gradually decreased, and some fine details (shown in the zoomed-in images) were also gradually blurred out with higher AFs.

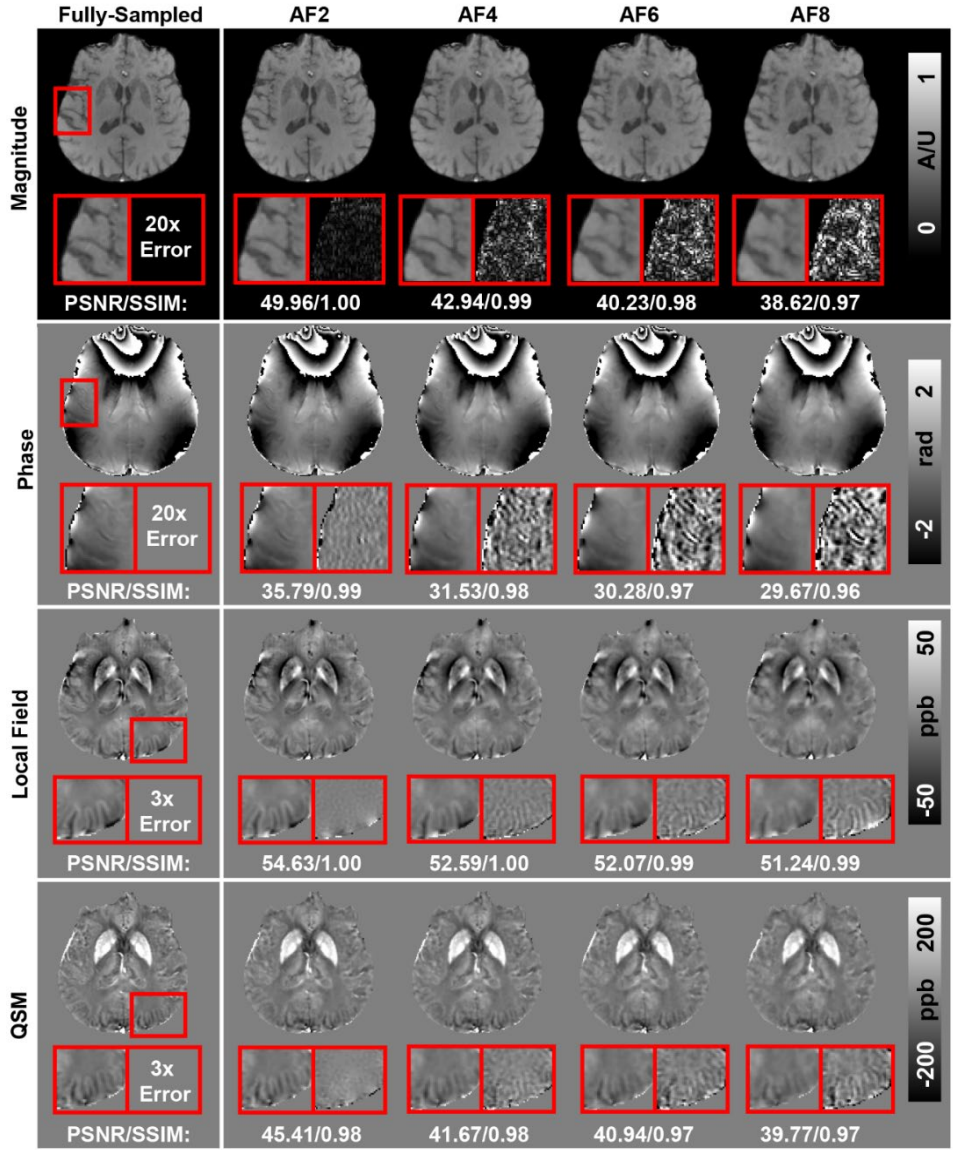

**Figure 10**. Performance evaluation of the proposed QSM accelerating framework against varying acceleration factors (AFs). Magnitude, phase, local field, and susceptibility maps are displayed from top to bottom rows, with corresponding PSNR and SSIM relative to the fully-sampled ground truth below the images. The first column shows the reconstruction of fully-sampled data, and the results of different AFs (from left to right: 2×, 4×, 6×, 8×) are displayed from the second column. Error maps from the zoomed-in regions are amplified to illustrate differences in structural details better.

## Discussion

In this work, we demonstrated the feasibility of using deep neural networks to accelerate the QSM acquisition. Specifically, we developed a DCRNet method to recover the GRE phase from the CS undersampled k-space acquisition, thus enabling the phase-based QSM acceleration. The proposed method was compared with two iterative algorithms (i.e., $CS_{PC}$, $CS_{PR}$) and one previous deep learning-based phase reconstruction method (i.e., Phase-Unet). The results on healthy volunteers and patients with multiple sclerosis and intracranial hemorrhage showed that the proposed deep learning QSM accelerating framework resulted in more accurate QSM reconstructions (substantially improved PSNR, SSIM, and susceptibility ROI measurements) with fewer artifacts and over-smoothness than conventional CS and the Phase-Unet approaches.

Although many previous deep neural networks [55-62] have been proposed for solving QSM reconstruction problems, most of them focused on solving the QSM dipole inversion; this work is the first to investigate how to use deep neural networks accelerating 3D QSM acquisition with incoherent subsampling strategies. To the best of our knowledge, only one previous work [17] studied how to accelerate QSM with the conventional iterative method $CS_{PC}$ on mouse brains. This paper compares the proposed method with existing literature on QSM CS-undersampling acceleration.

The deep learning-based DCRNet QSM accelerating framework has two main advantages over the iterative $CS_{PC}$ and $CS_{PR}$ methods. Firstly, it does not need to manually design feature representations (regularizations) based on top-down mathematical modelling since it can learn more effective ones from the training datasets and, therefore, potentially improve the reconstruction results. Secondly, the proposed method is much more computationally efficient than conventional iterative methods. Our test results on the human brain data confirmed that the proposed deep learning method substantially shortened the reconstruction time. For example, the reconstruction time of the proposed deep residual network is less than 1 minute, compared to 21 hours for $CS_{PC}$ and 28 hours for $CS_{PR}$ on a simulated single-channel 8-echo GRE brain volume of size 256×256×128×8(TEs), and is almost 10 thousand times faster than the $CS_{PC}$ method (about 80 hours) on the 32-channel in vivo ME-GRE subject (matrix size: 256×192×128×8(TEs)×32(receivers).

We also compared DCRNet with one previous deep learning-based MRI phase reconstruction method (Phase-Unet). It was found that the phase wraps in the reconstructed images were disrupted by Phase-Unet, making it problematic for the phase unwrapping process of the QSM pipeline. Our results showed that compared with DCRNet, Phase-Unet introduced severe artifacts near phase wraps in the field and susceptibility maps. In addition, the Mag-Unet and Phase-Unet reconstruct magnitude and phase images independently. On the other hand, the DCRNet method exploits the intrinsic relationship between magnitude and phase components through the complex convolution, leading to more accurate reconstructions for both.

The DCRNet neural network was trained on retrospectively undersampled GRE datasets from simulated single-channel MRI data with relatively high SNR. For the prospectively undersampled GRE acquisition in vivo, a 32-channel head coil was used for signal reception. The current DCRNet cannot directly reconstruct the individual channels due to the low SNR and receiver coil sensitivity. In this study, the coil sensitivities were estimated, and complex signals from individual channels were properly combined before de-aliasing with the DCRNet. Unlike the single-channel training datasets, the data consistency module was not feasible for the multi-channel combined images. It is observed that the lack of data consistency enforcement led to the effect of smoothness, particularly in the magnitude images (Fig. 9). To break through this limitation and make this network more generic to different receiver coil setups, we will develop deep learning models capable of directly reconstructing multiple-channel MRI data in the future. Additionally, ASSET acceleration and partial Fourier in the training datasets cause the intrinsic image resolution lower than 1 mm isotropic, which may also contribute to smoothness in the prospectively undersampled reconstruction results. Future studies will investigate the effects of mismatched image resolutions to improve DCRNet performance further.

The DCRNet method's performance as a function of the acceleration factor was also studied, and the results showed that no substantial errors or artifacts were introduced at all tested AFs (from 2× to 8×) using the proposed DCRNet framework. However, image quality gradually degraded with higher accelerations, showing increased blurring, which obscures structural detail. This study adopted deep neural networks for GRE phase undersampling reconstruction (i.e., DCRNet) and QSM dipole inversion (i.e., xQSM [49]). However, other QSM processing steps (e.g. phase unwrapping, background field removal) were still using traditional algorithms. Future work may develop a single or cascaded neural network that can directly reconstruct QSM from the undersampled phase acquisition. This strategy may minimize error propagation and amplification in each intermediate step and improve final results.

**Conclusion**

This study proposed a deep learning-based MRI phase accelerating framework, DCRNet, to accelerate data acquisition and reconstruction for QSM. The experimental results showed that the DCRNet method led to QSM images with fewer reconstruction errors and more accurate susceptibility measurements than the state-of-the-art iterative methods with the same accelerating factors. In addition, the DCRNet substantially reduced the reconstruction time by nearly 10 thousand times for each acquisition, from around 80 hours using iterative methods to only half a minute. The trained DCRNet model from healthy subjects can also be generalized to patients, as demonstrated in intracranial hemorrhage and multiple sclerosis cases.